\documentclass{aa}  
\usepackage{graphicx}
\usepackage{amsmath}
\usepackage{amssymb}
\usepackage{mathtools}
\usepackage{txfonts}
\newcommand\nuint{\nu_{\rm{i}}}
\newcommand\nuext{\nu_{\rm{e}}}

\newcommand\reff{R_{\rm{e}}}

\newcommand\Iz{I_0}
\newcommand\EulB{{\rm B}}
\newcommand\nua{\nu_{\rm a}}
\newcommand\Err{{\cal E}}
\newcommand\Errmax{{\cal E}^{\rm max}}
\newcommand\smax{s^{\rm max}}
\newcommand\st{s_{\rm t}}
\newcommand\rt{r_{\rm t}}
\newcommand\etamax{\eta^{\rm max}}
\newcommand\cu{c_1}
\newcommand\cd{c_2}

\begin{document} 

   \title{An accurate and simple, asymptotically matched deprojection of the S\'ersic law}

   \author{L. Ciotti\inst{1}\fnmsep\thanks{luca.ciotti@unibo.it}, L. De Deo\inst{1,2,3},
          \and
          S. Pellegrini\inst{1,2}
          }

   \institute{Department of Physics and Astronomy, University of Bologna, Via Gobetti 93/2, 40129 Bologna, Italy
             \and
             INAF -- Osservatorio di Astrofisica e Scienza dello Spazio di Bologna, Via Gobetti 93/3, 40129 Bologna, Italy
             \and
             International PhD College - Collegio Superiore, University of Bologna, Italy
             }

   \date{Submitted, October 17, 2024 - Resubmitted, January 2, 2025} 

 
  \abstract
   {The S\'ersic law reproduces very well the surface brightness profile of early-type galaxies, and therefore is routinely used in observational and theoretical works. Unfortunately, its deprojection can not be expressed in terms of elementary functions for generic values of the shape parameter $n$. Over the years, different families of approximate deprojection formulae have been proposed, generally based on fits of the numerical deprojection over some radial range.}
   {We searched for a very simple, accurate, and theoretically motivated deprojection formula of the S\'ersic law, without free parameters, not based on fits of the numerical deprojection, and holding for generic $n > 1$.}
   {The formula has been found by requiring it to reproduce the analytical expressions for the inner and outer asymptotic expansions of the deprojected S\'ersic law of given $n$, and by matching the two expansions at intermediate radii with the request that the total luminosity coincides with that of the original S\'ersic profile of same $n$.}
   {The resulting formula is algebraically very simple, by construction its inner and outer parts are the exact (asymptotic) deprojection of the S\'ersic law, and it depends on two coefficients that are analytical functions of $n$ of immediate evaluation. The accuracy of the formula over the whole radial range is very good and increases for increasing $n$, with {\it maximum} relative deviations from the true numerical deprojection of $\simeq 8\,10^{-3}$ for the de Vaucouleurs profile. In the Appendix, the extension of the proposed formula to profiles with $n <1$ is also presented and discussed. }
   {The formula obtained is a useful tool of simple use in the modeling of early-type galaxies. Its ellipsoidal generalization is immediate.}


   \maketitle
%
\section{Introduction}
The surface brightness profiles of early-type galaxies are described quite well by the empirical S\'ersic (1968) law (e.g., Caon, Capaccioli \& D'onofrio 1993), a generalization of the de Vaucouleurs (1948) profile. The S\'ersic law has also been used to model dwarf galaxies (e.g., Graham \& Guzman 2003; Battaglia et al. 2006), the bulges of spiral galaxies (Andredakis et al. 1995), and also globular clusters (e.g., Barmby et al. 2007). For this reason, the S\'ersic profile has been used intensively in observational works, and also investigated in depth from the theoretical point of view (e.g., see Graham \& Driver 2005, and references therein). For simplicity, in the following we restrict to spherical symmetry, but the generalization to the ellipsoidal case is not difficult (see Section 4; see also Chapter 11 in Ciotti 2021, hereafter C21). The S\'ersic profile of spherical galaxies can be written as
\begin{equation}
    I(R) =\Iz {\rm e}^{-b\, \eta^{1/n}},\quad \eta \equiv {R\over\reff},
\label{eq:Iser}
\end{equation}
where $n$ is the shape parameter, $\Iz$ the central surface brightness $\Iz$, $R$ is the radius in the projection plane, $\reff$ is the effective radius, i.e. the radius in the projection plane enclosing half of the total luminosity of the profile, and $b$ is a dimensionless parameter depending on $n$. The total luminosity associated with the profile in equation (\ref{eq:Iser}) is 
\begin{equation}
L = \Iz\reff^2 {2\pi n\over b^{2n}} \Gamma(2n),
\label{eq:Ltot}
\end{equation}
where $\Gamma(x)$ is the Euler's gamma function. Usually, the equation above is used to eliminate $\Iz$ in equation (\ref{eq:Iser}) in favour of $L$. As is well known the value of the parameter $b$ in equation (\ref{eq:Iser}) depends on $n$ via the transcendental equation $\Gamma(2n)/2 = \gamma(2n,b)$, where $\gamma(a,x)$ is the lower incomplete gamma function; the equation comes from the request that $\reff$ is the effective radius, and since $b(n)$ cannot be expressed in terms of elementary functions, several high-quality fits of the numerical solution have been presented in the literature (see e.g. Ciotti \& Bertin 1999, see also C21). Finally, Ciotti \& Bertin (1999) showed that $b(n)$ for $n\to\infty$ can be expressed as the asymptotic series
\begin{equation}
b = 2n - {1\over 3} + {4\over 405 n} + {46\over 25515 n^2}+O(n^{-3}),
\label{eq:bn}
\end{equation}
where higher order terms can be easily constructed. For practical purposes it should be noticed that the relative error (in absolute value) of equation (\ref{eq:bn}) truncated just to the first four terms compared to the true solution for $1\leq n \leq 10$ is already as small as $\simeq 3.8\times10^{-6}$ in the worst $n\simeq 1.2$ case, decreasing down to $\approx 10^{-8}$ for increasing $n$. Even for $n$ as low as $1/2$ the relative error is $\simeq 7\times10^{-4}$.

The deprojected luminosity density profile $\nu(r)$ of the S\'ersic law, the focus of the present paper, is given by the usual Abel deprojection integral (e.g., Binney \& Tremaine 2008, C21)
\begin{equation}
\nu(r) = -{1\over\pi} \int_r^\infty {d I(R)\over dR}{dR\over\sqrt{R^2 - r^2}},
\label{eq:Abel}
\end{equation}
where $r$ is the spherical radius. As is well known, $\nu(r)$ can not be expressed in terms of elementary functions for generic values of $n$. For this reason, a great effort has been devoted to build sufficiently accurate and manageable analytical approximations of the true (i.e., numerical) $\nu(r)$, usually valid over some more or less extended radial interval. We mention here a few of the main adopted approaches, with a list that is almost certainly not complete:

\begin{enumerate}

\item
{\it Purely numerical deprojections}: the $\nu(r)$ profiles are given as tables, in particular for the special case of the de Vaucouleurs $n=4$ law (Poveda et al. 1960, Young 1976).

\item
{\it Multi-gaussian deprojections}: the $\nu(r)$ profiles are determined from the property that the deprojection of a centered Gaussian is a Gaussian. The weights and dispersions of the Gaussian components adopted to expand $I(R)$ in equation (\ref{eq:Iser}) are determined numerically and given in tables. This approach was used for the $n=4$ profile by Bendinelli et al. (1993), and successively extended to generic S\'ersic profiles (e.g. Pechetti et al. 2020).

\item
{\it Single slope power-law fits}: in this important family $\nu(r)$ is approximated as 
\begin{equation}    
\nu(r)={L\over\reff^3}{b^{(3-\beta)n}\over 4\pi n\,\Gamma[(3-\beta)n]}s^{-\beta}{\rm e}^{-b\, s^{1/n}},\quad s\equiv{r\over\reff},
\label{eq:MM}
\end{equation}    
(Mellier \& Mathez 1987, Prugniel \& Simien 1997, see also Gerbal et al. 1997, Lima-Neto et al. 1999, Marquez 2000, Terzic \& Graham 2005). The expression for $\nu(r)$ in equation (\ref{eq:MM}) guarantees that for generic $\beta <3$ its total luminosity $L$ is the same as that of the S\'ersic profile of index $n$, effective radius $\reff$, and central surface brightness $\Iz$. The parameter $\beta$ is a function of $n$, and is obtained from a fit of the numerically deprojected profile over some prescribed radial range; in this way the value of $\beta$ also depends on the radial range used for the fit. Notice that $\reff$ is the effective radius of the true profile, not that of the deprojected approximated formula (even though they can be almost identical).

\item
{\it Refined single slope power-law fits}: $\nu(r)$ is reproduced with modifications of the single power-law fits, by considering more complicated functions than the exponential term in equation (\ref{eq:MM}), for example as a sum of exponentials (Simonneau \& Prada 2004). In this family we mention the very high-accuracy deprojected profile of Vitral \& Mamon (2020), obtained by fitting the residuals of equation (\ref{eq:MM}) with respect to the true profile, with bivariate polynomials of order 10 in terms of $\log(s)^i\log^j (n)$; the polynomials coefficients are given in numerical tables, and the lower limit of the range of $r/\reff$ was set to $10^{-3}$. In Vitral \& Mamon (2021) the lower limit was further reduced to $10^{-4}$.  

\item
{\it Special functions fits}: a high-quality fit of $\nu(r)$ in terms of the modified Bessel functions of second kind $K_{\nu}$, supplemented by tables for the numerical values of the parameters as a function of $n$, is given in Trujillo et al. (2002); a refinement of the formula was given in Emsellem \& van de Ven (2008). 

\item
{\it Analytical deprojections}: $\nu(r)$ is a Gaussian for $n=1/2$, and a modified Bessel function of the second kind $K_0$ for the exponential case $n=1$ (e.g., see Baes \& Ciotti 2019). Moreover, $\nu(r)$ can be expressed in terms of Meijer G functions for all rational values of $n$ (Mazure \& Capelato 2002), and in terms of Fox H functions for generic $n$ (Baes \& Gentile 2011, Baes \& Van Hese 2011). However these advanced transcendental functions are rarely used in common applications, and are not routinely implemented in numerical packages. In this family of analytically exact formulae, we also mention those based on the method of {\it asymptotic expansion} (Ciotti 1991, hereafter C91, see also C21) that give the asymptotic formulae for $\nu (r)$ for $r\to 0$ and for $r\to\infty$; as shown in Section 2, these formulae are the starting point of the present study.

\end{enumerate}

Nowadays, available computer algebra systems allow to integrate numerically equation (\ref{eq:Abel}) with arbitrarily large numerical precision, so that the need of sophisticated analytical approximations for $\nu(r)$ is not as compelling as it was when the de Vaucouleurs and S\'ersic laws were introduced. However, an analytical approximation of $\nu(r)$ can be still useful if it satisfies the following properties:

P1) Correct analytical behavior at the center and at large radii. In fact, one of the common uses of an analytical $\nu(r)$ is to obtain the radial trends of dynamical quantities (such as gravitational potential, solutions of Jeans equations, and so on) near the center, also in presence of a central supermassive black hole. Quite often the approximations of $\nu(r)$, obtained as fits over some finite radial range, are {\it not} correct at the center and/or at large radii (unless the correct trend is imposed to the functional form of the fitting function).

P2) Monotonically decreasing behavior for increasing $r$. This property of the S\'ersic law for $n\geq 1/2$ follows from the identity 
\begin{equation}
{d\,\nu(r)\over dr}=-{r\over\pi}\int_r^{\infty}{d\over dR}\left[{1\over R}{d I(R)\over dR}\right]{dR\over\sqrt{R^2-r^2}},
\end{equation}
established with integration by parts of equation (\ref{eq:Abel}), and then by differentiation with respect to $r$. For 
$n\geq 1/2$ it is easy to show that the integrand is nowhere negative, while for $0<n<1/2$ an asymptotic analysis shows that the profile is non-monotonic, with $d\,\nu(r)/dr > 0$ near the center, and negative at large radii.

P3) Conservation of the total luminosity. The proposed formula should have the same $L$ of the true S\'ersic profile of given $n$, $\reff$, and $\Iz$. It should be noticed that in general $\reff$ appearing in the approximate formulae is the effective radius of the true profile, not the effective radius of the approximated formula (even though, for high accuracy formulae the two values are very similar).

P4) Accuracy over a large/total radial range. The request is obvious, however for practical purposes differences of less than a few percents from the true $\nu(r)$ can be safely ignored in almost all observational and theoretical works. 

P5) Simplicity. As said, since a fast numerical deprojection with arbitrarily high accuracy is nowadays possible, only sufficiently simple formulae can be really useful in applications.  

The different formulae proposed so far satisfy some of the properties above, while are less satisfactory for some other. Here, we follow a different approach from that of previous works, i.e., {\it we do not use any fit}, but we build a formula starting from the analytically exact (asymptotic) deprojection of the S\'ersic law for large and small radii, restricting to $n>1$, the range of interest in applications. In this way, the proposed deprojection formula is correct by design for $r\to 0$ and $r\to\infty$. Then we match the two regimes introducing a {\it transition parameter} $p$, whose value is {\it uniquely} determined by $n$ and by the request that the total luminosity coincides with $L$ of the original S\'ersic law. The resulting formula is extremely simple and performs {\it extremely} well over the {\it entire} radial range, with relative errors of the order of $10^{-3}$ or less (with accuracy improving with increasing $n$). 

The paper is organized as follows. In Section 2 we construct and discuss the proposed deprojection formula, and in Section 3 the formula is compared to the true numerically deprojected S\'ersic profile. In Section 4 the main results are summarized.

\section{The asymptotically matched deprojection formula}
In our study we mainly focus on the deprojection $\nu(r)$ of the S\'ersic profile for $n>1$, the range encompassing the vast majority of cases of interest for early-type galaxies; in Appendix A however we consider in some detail the approximation of $\nu$ also for $n<1$ profiles. We do not consider the approximation of the deprojection of the $n=1$ profile because for this model $\nu$ can be written as a Bessel $K_0$ function: 
\begin{equation}
\nu(r) ={\Iz\over\reff}{b\over\pi}K_0(b s),
\label{eq:nuOne}
\end{equation}
where $s =r/\reff$, the convention adopted throughout the paper, and $b=b(1)\simeq 1.67835$. For future reference, we recall that near the origin
\begin{equation}
\nu(r)\sim {\Iz\over\reff}{b\over\pi}\left(\ln{2\over b s} - \gamma\right),\quad r\to 0,
\label{eq:nuOnecent}
\end{equation}
where $\gamma\simeq 0.57721$ is the Euler's constant, and the characteristic central logarithmic divergence of the deprojected exponential profile is apparent (see e.g. equation (A6) in C91, and equations 8.447.3 and 8.447.1 in Gradshteyn \& Ryzhik 2007). 

As shown in C91, for $0<n<1$ near the origin $\nu(r)$ converges to a finite value (dependent on $n$) as
\begin{equation}
\nu(r) \sim {\Iz\over\reff} {b^n \Gamma (1-n) \over \pi} {\rm e}^{-b\, s^{1/n}},\quad r\to 0.
\label{eq:nucent}
\end{equation}
For $n>1$ instead $\nu(r)$ near the origin diverges as\footnote{Due to a typo, the coefficient $2mb^{m-1}$ at the denominator of equation (9) in C91 and of equation (13.25) in C21 (for $m>1$), should be $2\pi m b^{-1}$. Notice that equation (A5) in C91, from which the correct equation (\ref{eq:nuint}) in this paper and the two mentioned equations in C91 and C21 derive, is correct. Again for a typo in equation (13.25) in C21 (for $m =1$), the $s^{1/2}$ at the denominator should be $s$, as in the correct equations (\ref{eq:nuOne}) in this paper and (A6) in C91.}
\begin{equation}
\nu(r) \sim \nuint (r)\equiv {\Iz\over\reff} \EulB\left({1\over 2}, {n-1\over 2n}\right) {b\over 2 \pi n}
{{\rm e}^{-b\, s^{1/n}}\over s^{1-1/n}},\quad r\to 0,
\label{eq:nuint}
\end{equation}
where $\EulB(x,y)$ is the Euler's beta function. In the external regions the leading term of the asymptotic expansion of $\nu(r)$ for a generic $n>0$ is finally given by
\begin{equation}
\nu(r) \sim\nuext (r)\equiv{\Iz\over\reff} \sqrt{{b\over 2\pi n}} 
{{\rm e}^{-b\, s^{1/n}}\over s^{1-1/(2n)}},\quad r\to\infty.
\label{eq:nuext}
\end{equation}
Of course, in equations (\ref{eq:nuOne})-(\ref{eq:nuext}) $\Iz$ can be expressed in terms of $L$ and $\reff$ using equation (\ref{eq:Ltot}).

These asymptotic expansions show that equation (\ref{eq:MM}) cannot be simultaneously correct at the center and in the external regions; therefore single power-law fits cannot be used to evaluate (rigorously) the dynamical properties of $\nu(r)$. However, the asymptotic expansions also explain why, after all, the single slope power-law fits in equation (\ref{eq:MM}) perform surprisingly well, with increasing accuracy for increasing $n$: in fact the algebraic factors in equations (\ref{eq:nuint})-({\ref{eq:nuext}) behave as $r^{1/n -1}$ at the center, and as $r^{1/(2n)-1}$ at large radii, so that for $n$ large enough the two exponents tend to the common limit of $-1$. For example, for $n=4$ the inner and outer power-law exponents for are already quite similar: $-0.75$ at the center, and $-0.875$ in the external regions (and in the Mellier \& Mathez 1987 formula the best-fit exponent is $\simeq -0.855$). Notice also how the asymptotic expansion is not uniform in terms of $n$, i.e., the central behavior of $\nu$ for $n=1$ cannot be obtained as the limit for $n\to 1^+$ of equation (\ref{eq:nuint}): it is to be expected that for $n$ near unity the deprojection fomulae based on equations (\ref{eq:nuint})-(\ref{eq:nuext}) will be (slightly) less accurate than for (say) $n\geq 2$.  

Thanks to equations (\ref{eq:nuint})-(\ref{eq:nuext}) we can now proceed to the construction of a high-accuracy approximated deprojected S\'ersic profile that is {\it not} a fit, with no free parameters, monotonically decreasing, with the same total luminosity and the same exact trend in the internal and external regions as the true deprojected profile. The idea is to exploit the fact that, for $n>1$, the ratio $\nuint(r)/\nuext(r)\propto r^{1/(2n)}$ is a monotonically increasing function, vanishing at the origin and diverging for $r\to\infty$. This allows to satisfy all the mentioned requirements (except the conservation of $L$, that we discuss later on) with the elementary trial function  
\begin{equation}
\nua(r)\equiv\displaystyle{{\nuint(r)\over 1+ \nuint(r)/\nuext(r)}}=\displaystyle{{\nuext(r)\over 1+ \nuext(r)/\nuint(r)}}.
\label{eq:testnu}
\end{equation}
Notice that the request of exact central and external behaviors (i.e., that $\lim_{r\to 0} \nua/\nuint =1$, and $\lim_{r\to\infty}\nua/\nuext =1$) implies that the coefficients in front of  $\nuint(r)$ and $\nuext(r)$ in equation (\ref{eq:testnu}) must be unity, as must also be the additive constant in its denominator.
Notice also that the proposed $\nua$ approaches both $\nuint$ and $\nuext$ from below.

Of course, $\nua(r)$ cannot be expected to describe well the true profile in the intermediate radial region; for example volume integration of  $\nua(r)$ gives a total luminosity that {\it underestimates} $L$ of the associated S\'ersic profile by $\simeq 19\%$ for $n=2$, $\simeq 24\%$ for $n=4$, and $\simeq 27\%$ for $n=10$. As a consequence, while preserving all its positive features, we must modify equation (\ref{eq:testnu}) so that its total luminosity coincides with $L$. The idea is to introduce a matching parameter $p$ (dependent on $n$) to be {\it uniquely} determined by the conservation of $L$. The most natural choice is to modify equation (\ref{eq:testnu}) as 
\begin{equation}
\nua(r) = \displaystyle{\nuint(r)\over\left\{1 +\left[\nuint(r)/\nuext(r)\right]^p\right\}^{1/p}}
={L\over\reff^3}{\cu\,{\rm e}^{-b\, s^{1/n}}\over \left [1 + \cd^p\, s^{p/(2n)}\right]^{1/p}s^{1-1/n}},
\label{eq:nua}
\end{equation}
where $b$ is again given in equation (\ref{eq:bn}), the second formula is expressed in terms of $L$ and $\reff$ of the true S\'ersic profile, and from equations (\ref{eq:nuint})-(\ref{eq:nuext}) the two coefficients are the analytical functions 
\begin{equation}
\cu\equiv {b^{2 n+1}\over 4 \pi ^2 n^2 \Gamma (2 n)}\, \EulB\left({1\over 2}, {n-1\over 2n}\right),\quad 
\cd\equiv\sqrt{{b\over 2 \pi  n}}\,\EulB\left({1\over 2}, {n-1\over 2n}\right).
\label{eq:coeff}
\end{equation}
For $p>0$ the asymptotic properties of equation (\ref{eq:testnu}) are retained, in agreement with the requests in P1, P2, P5 in the Introduction. Moreover, for $p>1$ it follows that $1+x\geq (1+x^p)^{1/p}$ $\forall x\geq 0$ (while the inequality inverts for $0<p <1$), so that for $p>1$ the denominator of equation (\ref{eq:nua}) stays below that of equation (\ref{eq:testnu}), correcting the total luminosity deficiency of the latter. Summarizing, the value of $p$ is uniquely determined by the request that equation (\ref{eq:nua}) satisfies P3, i.e. by the condition
\begin{equation}
4\pi\int_0^{\infty}\nua(r)r^2dr = L.
\label{eq:P3}
\end{equation}
We numerically determined the values $p(n)$ that lead to the luminosity conservation for a selection of representative values of $n$; they are plotted in Figure \ref{fig:plot_p} and given in Table \ref{table:pmore}. Note how $p(n)$ is greater than 1 for $n\gtrsim$, as expected, and how it increases for increasing $n$, converging to the limit\footnote{Volume integration of equation (\ref{eq:nua}) in the limit of $n\to\infty$ can be obtained by asymptotic methods noticing that the denominator of $\nua$ tends to the constant $(1+\pi^{p/2})^{1/p}$, so that the limit value of $p$ is given by the solution of $1+\pi^{p/2}=2^p$, in agreement with the value determined from the numerical integration described in the text.} of $p\simeq 2.13645$ for very large $n$. For $n\to 1$ the value of $p(n)$ drops below unity, and in fact in Appendix A we prove that $p(1)=0$; decreasing further $n$, the asymptotically matched deprojection $\nua$ changes as in equation (\ref{eq:nualessone}), and the function $p(n)$ now increases.

\begin{figure}
    \centering
    \includegraphics[width=\linewidth]{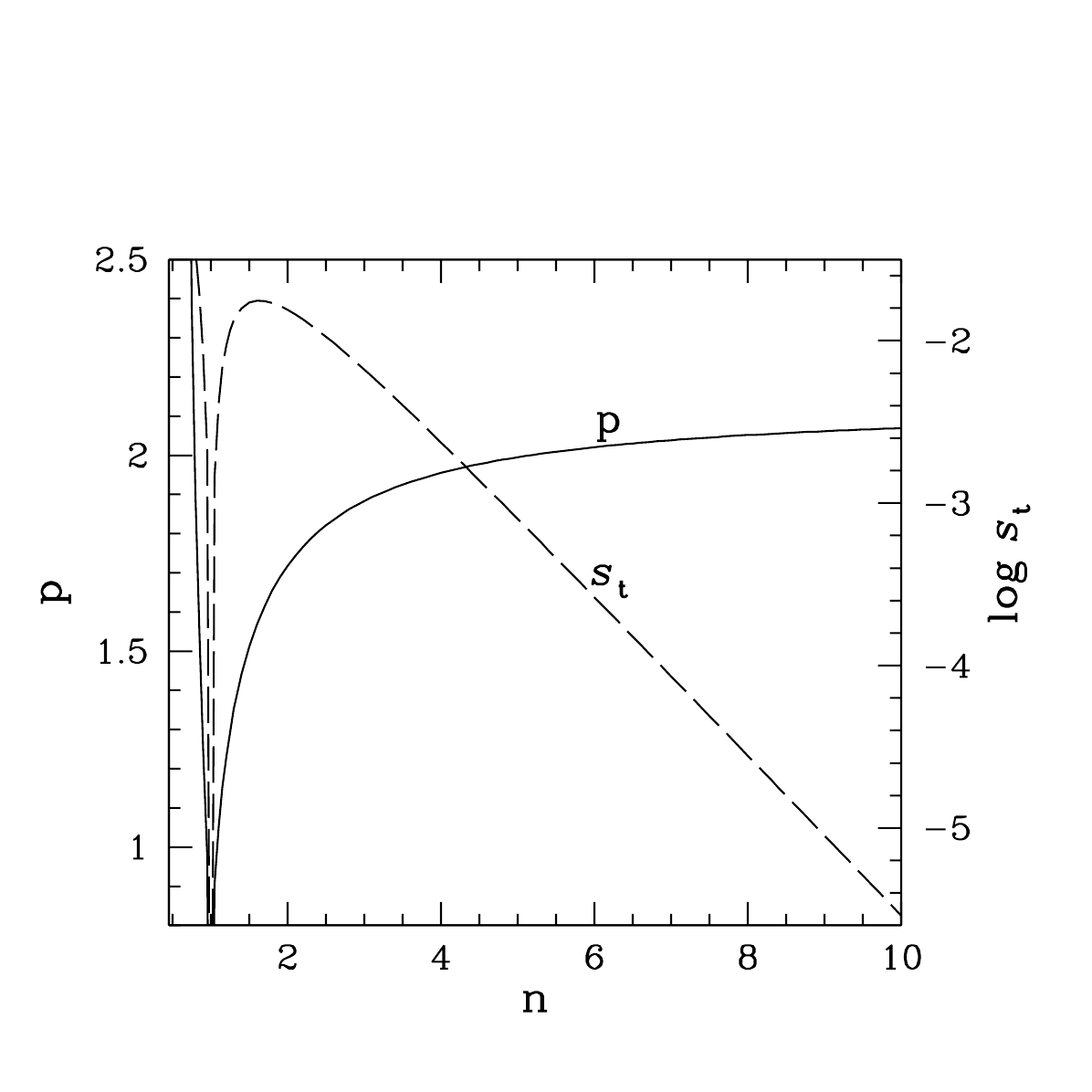}
    \vskip -0.5 truecm
    \caption{The matching parameter $p$ as a function of the S\'ersic index $n$ (solid line), determined solving numerically equation (\ref{eq:P3}), so that the total luminosity of the profile $\nua(r)$ in equation (\ref{eq:nua}) coincides with $L$ of the true profile in equation (\ref{eq:Iser}). As explained in Appendix A, $p(n)$ vanishes for $n\to 1$ and then increases for decreasing $n<1$ (see Table \ref{table:pmore}). The dashed line gives the transition radius (in units of $\reff$) as defined in equation (\ref{eq:rtrans}): for $n <1$, $\st$ increases for decreasing $n$, remaining however small.}
    \label{fig:plot_p}
\end{figure}

It is also useful to define the {\it transition radius} $\rt$ as the radius at which $\nuint(\rt)=\nuext(\rt)$, i.e. from equation (\ref{eq:nua})
\begin{equation}
\st ={\rt\over\reff} = {1\over\cd^{2n}}.
\label{eq:rtrans}
\end{equation}
Qualitatively, for distances from the center smaller than $\st$ the profile is dominated by $\nuint(r)$, while for larger radii the profile is better and better described by $\nuext(r)$. In Figure \ref{fig:plot_p} the dashed line shows the function $\st (n)$; notice the non-monotonicity at low values of $n\lesssim 2$, with a maximum of $\st\simeq 0.02$ reached at $n\simeq 1.63$, and the following decrease to $\st(4)\simeq 0.002$, $\st(6)\simeq 0.0002$, and so on, with $\st\to 0$ for $n\to 1$. These trends can be explained with a careful asymptotic analysis of equation (\ref{eq:rtrans}), that shows that $\st\sim \pi (n-1)^2/[2b(1)]$ for $n\to 1^+$, and $\st\sim {\rm e}^{1/6}/(4\pi^n)$ for $n\to \infty$. In practice $\st$ is always very small, an indication that $\nu(r)$ is almost completely dominated by the external asymptotic profile. For sake of completeness, we notice that for $n <1$, $\st$ instead increases for decreasing $n$, as illustrated in Figure \ref{fig:plot_p} and proved in Appendix A.}
 
\section{Results}
\label{sec:num_sol}
We checked the accuracy of $\nua$ in equation (\ref{eq:nua}) by computing the relative error profile 
\begin{equation}
\Err (r)\equiv{\nua(r) - \nu(r)\over\nu(r)},
\label{eq:nuerr}
\end{equation}
where $\nu(r)$ is the numerically deprojected profile. As $\nu(r)$ diverges at the center for $n\geq 1$, a poor treatment of the integral in equation (\ref{eq:Abel}) can easily lead to a significant accuracy loss for small radii, affecting in turn the robustness of the indicator $\Err (r)$. In order to control the deprojection, we started from equation (7) of C91 (which holds for generic $n>0$), written in term of the reduced radius $\alpha \equiv b s^{1/n}$, where the central divergence of $\nu(r)$ for $n>1$ is factored out exactly:
\begin{equation}
\begin{split}
    \nu(r) &= {\Iz\over\reff}{b^n \alpha^{1-n}\over\pi} \int_1^{\infty} {{\rm e}^{-\alpha\, t} dt\over\sqrt{t^{2n} - 1}}\\
             &= {\Iz\over\reff} {b^n \alpha^{1-n} {\rm e}^{-\alpha}\over n \pi} \int_0^\infty 
                  {{\rm e}^{-\alpha\, [\left(x^2 + 1\right)^{1/(2n)}-1]}\over 
                    \left(x^2 + 1\right)^{1 - 1/(2n)}} dx.
\label{eq:num_dep}
\end{split}
\end{equation}
In this way the resulting integral is now convergent. In the second line we adopted the substitution $x=\sqrt{t^{2n}-1}$, so that the new integrand is free of the singularity, improving further the numerical stability. Motivated by the functional form of equations (\ref{eq:nuint})-(\ref{eq:nuext}), we finally factored out the exponential term ${\rm e}^{-\alpha}$.

\begin{table}
\caption{Maximum relative errors} 
\label{table:p}     
\centering                          
\begin{tabular}{ c  c  c  c  c }       
\hline\hline               
\rule{0pt}{2.5ex} $n$ & $\Errmax$ & $\log\smax$ & $\Delta\mu^{\rm max}$ & $\log\etamax$ \\[2.5pt] 
    (1) & (2) & (3) & (4) & (5)\\
\hline
   1.5 & 0.089 & $-2.616$ & $-0.047$ & $-3.778$\\      
   2.0 & 0.033 & $-2.591$ & $-0.024$ & $-3.499$\\
   2.5 & 0.018 & $-2.789$ & $-0.015$ & $-3.624$\\
   3.0 & 0.012 & $-3.071$ & $-0.011$ & $-3.911$\\
   3.5 & 0.009 & $-3.445$ & $-0.009$ & $-4.000$\\
   4.0 & 0.008 & $-3.854$ & $-0.007$ & $-4.000$\\
   4.5 & 0.007 & $-4.000$ & $-0.006$ & $-4.000$\\
   5.0 & 0.006 & $-4.000$ & $-0.005$ & $-4.000$\\
   5.5 & 0.005 & $-4.000$ & $-0.005$ & $-4.000$\\
   6.0 & 0.004 & $-4.000$ & $-0.004$ & $-4.000$\\
   6.5 & 0.004 & $-4.000$ & $-0.003$ & $-4.000$\\
   7.0 & 0.003 & $-4.000$ & $-0.003$ & $-4.000$\\
   7.5 & 0.003 & $-4.000$ & $-0.003$ & $-4.000$\\
   8.0 & 0.003 & $-4.000$ & $-0.003$ & $-4.000$\\
   8.5 & 0.002 & $-4.000$ & $-0.002$ & $-4.000$\\
   9.0 & 0.002 & $-4.000$ & $-0.002$ & $-4.000$\\
   9.5 & 0.002 & $-4.000$ & $-0.002$ & $-4.000$\\
   10  & 0.002 & $-4.000$ & $-0.002$ & $-4.000$\\
\hline                               
\end{tabular}
\tablefoot{Column (2): largest relative deviation of $\nua(r)$ in equation (\ref{eq:nua}) compared to the true, numerically deprojected $\nu(r)$  in equation (\ref{eq:num_dep}), computed from $\Err(r)$ in equation (\ref{eq:nuerr}). Column (3): position  $\smax$ of the largest deviation $\Errmax$. When $\smax <10^{-4}$, the deviation at $s=r/\reff =10^{-4}$ is reported. Column (4): largest surface brightness difference (in magnitudes) between the true S\'ersic profile and the projection of $\nua(r)$,  computed from $\Delta\mu (R)$ in equation (\ref{eq:Delmu}). Column (5): position $\etamax$ of the largest magnitude difference $\Delta\mu^{\rm max}$. When $\etamax <10^{-4}$, the magnitude deviation at $\eta=R/\reff = 10^{-4}$ is given.}   
\end{table}

\begin{figure}
    \centering
    \includegraphics[width = \linewidth]{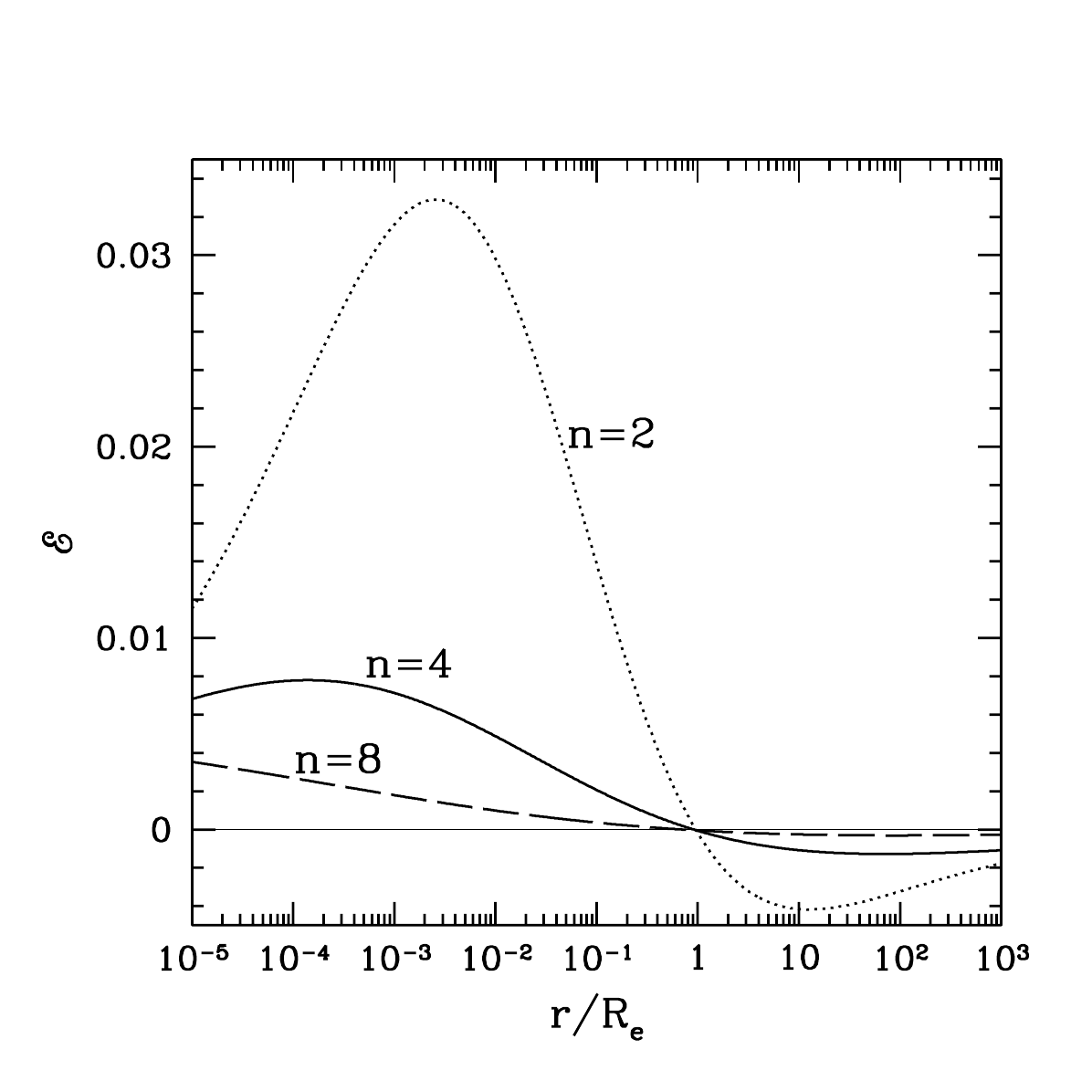}
        \vskip -0.5 truecm
    \caption{Radial trend of the relative error in equation (\ref{eq:nuerr}) between the asymptotically matched and the numerically deprojected profiles for $n=2,$ 4, and 8.}
    \label{fig:nuerr}
\end{figure}

The results are surprisingly good, and qualitatively similar for all values of $n$, with accuracy increasing for increasing $n$. The profile of $\Err(r)$ is shown in Figure \ref{fig:nuerr} for the three representative cases of $n=2$, $4$, and $8$; the largest error occurs at small radii. In general $\nua(r)$ tends to overestimate the true $\nu(r)$ at small radii, before converging to it (by construction) for $r\to 0$. Similarly, at large radii $\nua(r)$ tends to underestimate the true profile before converging to it (by construction) for $r\to\infty$. This behavior is due to the fact that while $\nua(r)$ lies below $\nuint(r)$ and $\nuext(r)$, the leading terms $\nuint(r)$ and $\nuext(r)$ remain above the true $\nu (r)$: in the central regions $\nu(r)<\nua(r)<\nuint(r)$, and in the external parts $\nua(r)<\nu(r)<\nuext(r)$. Of course, for $r\to 0$ and $r\to\infty$, the two inequalities become equalities.

Table \ref{table:p} reports (with its sign) the maximum relative deviation of $\Err(r)$ over the whole radial range, and the radius $\smax$ at which the maximum deviation is reached. As also shown in Figure \ref{fig:nuerr}, $\smax$ moves towards the center and $\Errmax$ decreases for increasing $n$. When $s^{\rm max}<10^{-4}$ (corresponding to models with $n\gtrsim 4.2$), i.e., the largest deviation is reached at radii where a continuous stellar density profile lacks astrophysical meaning, in the $\Errmax$ column we just report the value of $\Err$ at $s=10^{-4}$; in general this value is the maximum deviation for all $s > 10^{-4}$, due to the shape of $\Err (r)$. 

\begin{figure}
    \centering
    \includegraphics[width = \linewidth]{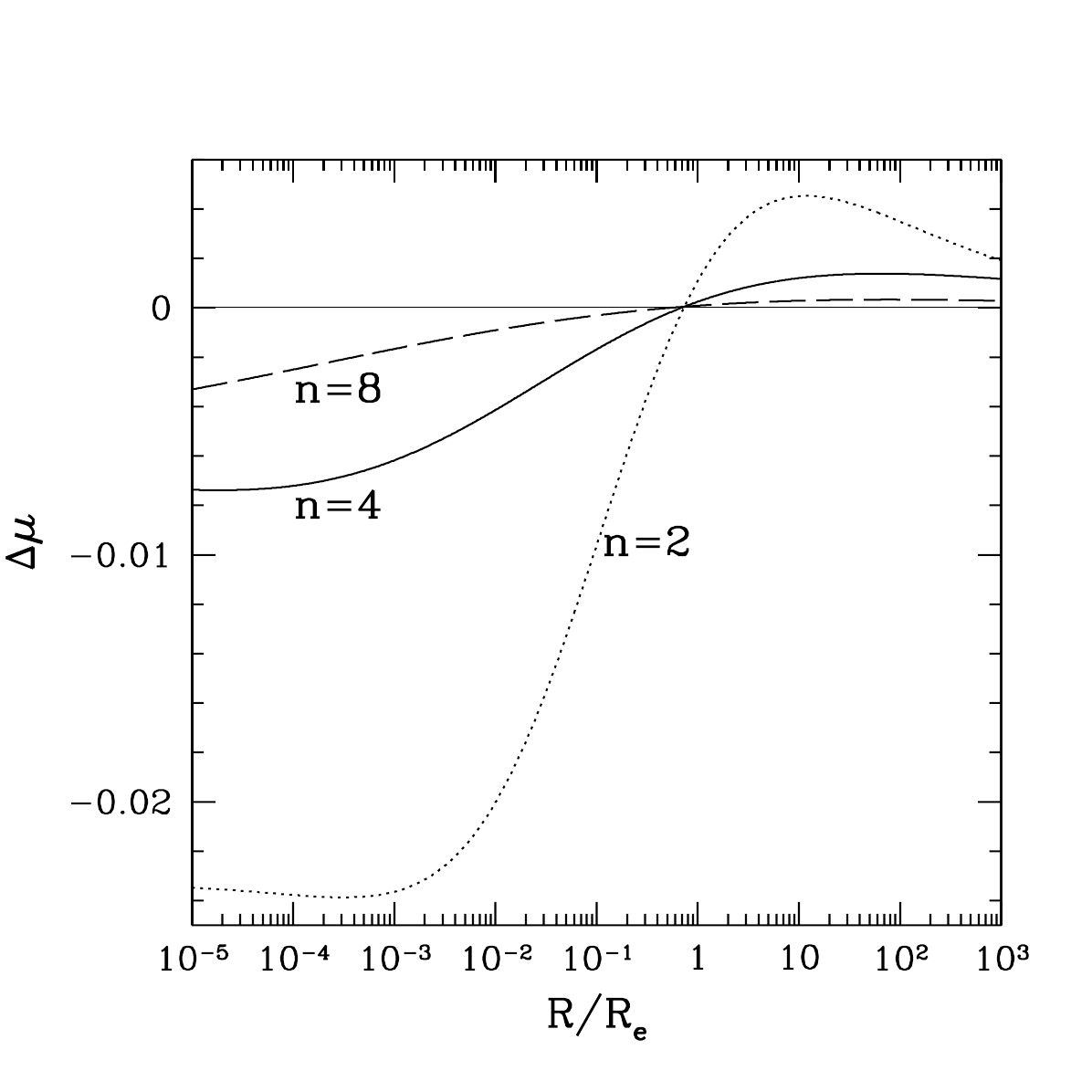}
    \vskip -0.5 truecm
    \caption{Difference in magnitude associated with the projected profile obtained from the asymptotically matched density $\nua(r)$ in equation (\ref{eq:nua}), and the true S\'ersic profiles with $n = 2$, 4, and 8.}
    \label{fig:muerr}
\end{figure}

To quantify the deviation of $\nua(r)$ from the true profile, we also consider the radial trend of the difference (in magnitudes) between the S\'ersic profile in equation (\ref{eq:Iser}), and the projected profile $I_{\rm a}(R)$ obtained from $\nua(r)$, i.e. we consider 
\begin{equation}
\Delta\mu (R)\equiv -2.5 \log {I_{\rm a}(R)\over I(R)}.
\label{eq:Delmu}
\end{equation}
Figure \ref{fig:muerr} shows $\Delta\mu (R)$ for the $n=2$, 4, and 8 profiles; the behavior is qualitatively similar for all $n>1$. As expected, the errors are extremely small, and decrease for increasing $n$, confirming that the accuracy of $\nua(r)$ is such that it can certainly be used for all practical purposes. In Columns (4) and (5) of Table \ref{table:p} we report the maximum magnitude difference, and the radius at which it is reached, for representative values of $n$; again, when the maximum deviation is reached inside $s=10^{-4}$ we give the magnitude difference at $s=10^{-4}$. 

Of course, it is particularly important that the errors compare well with those obtained from other high-accuracy fitting methods (e.g., Vitral \& Mamon 2020), but are now associated with a mathematically motivated, algebrically simple, and (except for the $p$ value) fully analytical formula; we stress again that this formula is not the result of a fit. 

\section{Conclusions}
The S\'ersic law is the standard profile to describe the surface brightness of early-type galaxies and other astrophysical systems. Unfortunately, its deprojection $\nu(r)$  can not be expressed in terms of elementary functions for generic values of the shape parameter $n$. Even though nowadays thanks to powerful computer algebra systems a numerical deprojection can be performed very efficiently and with arbitrarily high accuracy, in some applications the availability of simple and manageable formulae can be useful, and in fact over the years several analytical approximations of $\nu(r)$ have been proposed. Some of these formulae are very accurate, but in general they are  based on "ad hoc" fits of the numerical $\nu(r)$ deprojection over some radial range, therefore missing -- with some exceptions -- the correct asymptotic behavior of $\nu(r)$ at small and large radii; also, they may require the input of several numerically determined coefficients, sometimes given as tables.

In this work, we followed a different approach, and we constructed a simple algebraic formula $\nua(r)$ (not more complicated than a rational function) to approximate the true $\nu(r)$ for $n>1$. The derived formula is not a fitting formula, it contains only two analytical coefficients that depend on $n$, and came from requiring $\nua(r)$ to be asymptotically exact at small and large radii, and to reproduce the monotonic decrease of $\nu(r)$ with $r$ when $n\geq 1/2$. The two asymptotic regimes are joined with the introduction of a matching parameter $p(n)$, with a value {\it uniquely} determined by the request that the total luminosity of $\nua(r)$ is the same as that of the original S\'ersic profile of same index $n$, effective radius $\reff$ and central surface brightness $\Iz$. A grid of values of $p(n)$ is provided in Table \ref{table:pmore}, for $0.55\leq n\leq 11.6$; further values of $p(n)$ can be easily computed by interpolation or numerical integration of equation (\ref{eq:P3}). It is also possible to produce high-accuracy analytical interpolation formulae for $p(n)$, but we do not pursue this possibility in this paper. We tested $\nua(r)$ against the true $\nu(r)$, and we found its accuracy impressive at all radii, especially considering its extreme simplicity. The {\it maximum} absolute value of the relative difference between $\nu(r)$ and $\nua(r)$ decreases for increasing $n$, and is of the order of $\simeq 10^{-3}$ for $n\gtrsim 4$; for example, for the $R^{1/4}$ law, in the range $0.1\reff < R<10\reff$ relative errors are smaller than $0.002$. The accuracy of the formula is also tested by projecting $\nua(r)$ and then comparing the surface brightness profile with that of the S\'ersic law, expressed in magnitudes: the maximum absolute value of the relative difference is very small, smaller than $0.05$ mag in the worst low-$n$ case reported, decreasing to $0.007$ mag for the de Vaucouleurs profile and to $0.002$ mag or less for larger values of $n$. Due to its simplicity, high accuracy, and the fact that its properties came from imposing a match with the known properties of the true deprojected S\'ersic law, we believe the $\nua(r)$ in equation ({\ref{eq:nua}) is the best (theoretically motivated) compromise currently available between simplicity and accuracy. 


We finally notice that the generalization of the formula to axisymmetric and triaxial ellipsoidal systems is immediate. In fact, in order to describe a S\'ersic triaxial system of total luminosity $L$, stratified on 
\begin{equation}
m^2={x^2\over a^2}+{y^2\over b^2}+{z^2\over c^2},\quad q_y \equiv {b\over a},\quad q_z \equiv {c\over a},
\end{equation}
with $a\geq b\geq c$, in the expression for $\nua(r)$ in equation (\ref{eq:nua}) one substitutes the scaled radius $s$ with $m$, $\reff$ with $a$, and divides the resulting formula by the product $q_y q_z$; in this way the total luminosity of the resulting $\nua(m)$ is independent of the flattening. The projection of the density $\nua$ of such triaxial system at any inclination of the line-of-sight can be done as described in Chapter 11 in C21; in particular, it turns out that $a$ is the semimajor axis of the effective isophotal ellipse when projecting $\nua(m)$ along the intermediate ($y$) or short ($z)$ axis, and its value coincides with $\reff$ of the spherical limit of $\nua(m)$. This ellipsoidal generalization of $\nua(r)$ could be a useful improvement for the modeling technique based on triaxial S\'ersic ellipsoids presented in van de Ven \& van der Wel (2021).

\begin{acknowledgements}
We thank the Referee, Gary Mamon, for a very careful reading of the manuscript, insightful comments and useful suggestions. LDD thanks Arjen van der Wel for useful discussions.
\end{acknowledgements}

%
%

\appendix
\onecolumn
\section{The asymptotically matched formula for $n\leq 1$}

The paper is mainly focussed on the case of S\'ersic profiles with $n >1$, however it is of interest to illustrate how the proposed approach works for the cases $n\leq 1$.

We first show that the pointwise limit of equation (\ref{eq:nua}) for $n\to 1^+$ and {\it an arbitrary but fixed} $p>0$ converges to the profile
\begin{equation}
\nua(r) ={L\over\reff^3}{b(1)^{5/2}\over (2\pi)^{3/2}}{{\rm e}^{-b(1)\, s}\over\sqrt{s}}.
\label{eq:nualim}
\end{equation}
In fact the coefficients $c_1$ and $c_2$ in equation (\ref{eq:coeff}) both diverge for $n\to 1$, due to the divergence of the beta function as $2/(n-1)$; in particular, the divergence of $c_2$ implies that for arbitrary but fixed $p>0$, $\nua$ in equation (\ref{eq:nua}) converges everywhere to the profile $\nuext$ of the $n=1$ deprojected density in equation (\ref{eq:nuOne}). The volume integral of $\nua(r)$ in equation (\ref{eq:nualim}) gives a total luminosity of $3L/(2\sqrt{2})\simeq 1.06 L$. Therefore, for a generic but fixed $p>0$, the total luminosity of the pointwise limit of $\nua$ in equation (\ref{eq:nua}) for $n\to 1$ does not converge to the total luminosity of the true $n=1$ profile (even if it is quite close to it). Of course, this behavior is to be expected, as the limit of $\nua$ for $n\to 1$ should be taken {\it not} at a fixed $p$, but along the sequence $p(n)$ defined implicitly by equation (\ref{eq:P3}), so that the conservation of total $L$ is guaranteed. It follows immediately that $\lim_{n\to 1^+}p(n)=0$, because for any strictly positive limit value $p(1)>0$, equation  (\ref{eq:nualim}) would be reobtained, violating the total luminosity conservation. In fact, $p(n)$ is numerically found already below unity for $n=1.05$ (see Table \ref{table:pmore}), with values monotonically decreasing for decreasing $n$ towards unity (see Table \ref{table:pmore}). 

We now consider the $n<1$ profiles. From equations (\ref{eq:nucent}) and (\ref{eq:nuext}) it follows that for $1/2<n<1$ again $\nuint/\nuext\to\infty$ for $s\to\infty$, and $\nuint/\nuext\to 0$ for $s\to 0$, so that the arguments used to build equation (\ref{eq:nua}) still apply, and then
\begin{equation}
\nua(r) = \displaystyle{\nuint(r)\over\left\{1 +\left[\nuint(r)/\nuext(r)\right]^p\right\}^{1/p}}
={L\over\reff^3}{\cu\,{\rm e}^{-b\, s^{1/n}}\over \left [1 + \cd^p\, s^{p-p/(2n)}\right]^{1/p}},
\label{eq:nualessone}
\end{equation}
where
\begin{equation}
\cu\equiv {b^{3n}\Gamma (1-n)\over 2\pi^2 n\Gamma(2n)},\quad 
\cd\equiv\sqrt{{2n\over\pi}}b^{n-1/2}\Gamma(1-n).
\label{eq:coefflo}
\end{equation}

For $n=1/2$ the deprojected S\'ersic profile is Gaussian 
\begin{equation}
\nu(r)={I_0\over\reff}\sqrt{{b\over\pi}}{\rm e}^{-b s^2}={L\over\reff^3}\left({b\over\pi}\right)^{3/2}{\rm e}^{-b s^2},
\label{eq:gauss}
\end{equation}
where $b=b(1/2)=\ln 2$, and $\nu(r)=\nuint(r)=\nuext(r)$, i.e. for $n=1/2$ the true deprojected profile coincides at all radii with the asymptotic profiles in equations (\ref{eq:nucent}) and (\ref{eq:nuext}). It follows that $\nuint(r)/\nuext(r) =1$ everywhere, and then equations (\ref{eq:nualessone}) and (\ref{eq:P3}) require that $\lim_{n\to 1/2^+}p(n)=\infty$.  This is nicely confirmed by the numerically recovered values of $p$ (see Table \ref{table:pmore}).

As for the $n>1$, it is natural to define the transition radius $\rt$, where $\nuint(\rt)=\nuext(\rt)$, and from equation (\ref{eq:nualessone}) we obtain the analogous of equation (\ref{eq:rtrans})
\begin{equation}
\st =\cd^{{2n\over 1 -2n}}.
\label{eq:rtranslessone}
\end{equation}
Again, a careful asymptotic analysis shows that $\st\sim\pi (n-1)^2/[2b(1)]$ for $n\to 1^-$, while $\lim_{n\to 1/2^+}\st ={\rm e}^{-(1+\gamma)/2}/(2\sqrt{\ln 2})\simeq 0.273$, in agreement with Figure \ref{fig:plot_p}: notice however that for $n=1/2$ the concept of transition radius loses its meaning, because $\nuint(r)=\nuext(r)$ everywhere.

For sake of completeness, we also consider the range $0<n<1/2$, where the situation changes. In fact, now $\nuint/\nuext\to 0$ for $s\to\infty$ and diverges for $s\to 0$, so that equation (\ref{eq:nualessone}) with a positive $p$ would fail to reproduce the correct asymptotic trends of the true profile. However, it is not difficult to show that $\nua(r)$ in equation (\ref{eq:nualessone}) with {\it negative} values $p$ recovers the asymptotic trends of the true deprojected $\nu(r)$. 

In analogy with equation (\ref{eq:num_dep}), we finally give the integral expression of $\nu(r)$ to be used for high-accuracy deprojection when $n<1$
\begin{equation}
\nu(r) = {\Iz\over\reff} 
            {b^n {\rm e}^{-\alpha}\over n\pi} \int_0^\infty 
            {
            {\rm e}^{-[\left(x^2 + \alpha^{2n}\right)^{1/(2n)}-\alpha]}
            \over 
            \left(x^2 + \alpha^{2n}\right)^{1 - 1/(2n)}
            } dx,
\label{eq:num_deplo}
\end{equation}
obtained with the change of variable $x=t/\alpha^n$ in equation (\ref{eq:num_dep}). Notice that equation (\ref{eq:num_deplo}) avoids the problem of equation (\ref{eq:num_dep}) when used with $n<1$, namely the fact that the factor $\alpha^{1-n}$ therein would vanish for $s\to 0$, and the integral would diverge, as obvious from the fact that $\nu(0)$ is finite for $n<1$.

\section{Numerically determined values of $p(n)$}

Table \ref{table:pmore} gives the values of the matching parameter $p$ in equation (\ref{eq:nua}) for a grid of values of $n$. Two decimal digits of $p(n)$ are sufficient to guarantee that the total luminosity of $\nua(r)$ is the same as that of the Sersic profile of given $\Iz$, $\reff$ and $n$, with relative differences smaller than $10^{-3}$ for all values of $n$ reported; with three decimal digits the differences become much smaller than $10^{-4}$. For values of $n$ not reported in the Table, $p(n)$ can be easily computed by numerical integration of equation (\ref{eq:P3}), or by numerical interpolation of the values provided.

\begin{table}[h!]         
\caption{A grid of values of the matching parameter $p(n)$}
\label{table:pmore}     
\centering                          
\begin{tabular}{ c  c | c  c | c  c | c  c | c  c | c  c | c  c}       
\hline\hline              
\rule{0pt}{2.5ex} $n$ & $p$ & $n$ & $p$ & $n$ & $p$ & $n$ & $p$ & $n$ & $p$ & $n$ & $p$ & $n$ & $p$\\[2.5pt] 
\hline
0.55 &13.19 & 1.5	&1.511&   3.2   &1.902& 4.9	&1.992& 6.6	&2.032& 8.3	&2.055 & 10.0 & 2.069\\
0.60 &6.479 & 1.6	&1.568&   3.3	&1.910& 5.0	&1.995& 6.7	&2.034& 8.4	&2.056 & 10.1 & 2.070\\ 
0.65 &4.224 & 1.7	&1.615&   3.4	&1.918& 5.1	&1.998& 6.8	&2.035& 8.5	&2.057 & 10.2 & 2.071\\
0.70 &3.082 & 1.8     &1.654&   3.5	&1.925& 5.2	&2.001& 6.9     &2.037& 8.6	&2.058 & 10.3 & 2.071\\
0.75 &2.383 & 1.9	&1.688&   3.6	&1.932& 5.3	&2.004& 7.0	&2.038& 8.7	&2.059 & 10.4 & 2.072\\
0.80 &1.903 & 2.0	&1.718&   3.7	&1.938& 5.4	&2.007&7.1	&2.040& 8.8	&2.060 & 10.5 & 2.073\\
0.85 &1.542 & 2.1	&1.743&   3.8	&1.944& 5.5	&2.009& 7.2	&2.041& 8.9	&2.061 & 10.6 & 2.073\\
0.90 &1.248 & 2.2	&1.766&   3.9	&1.950& 5.6	&2.012&7.3	&2.043& 9.0	&2.061 & 10.7 & 2.074\\
0.95 &0.974 & 2.3	&1.786&   4.0	&1.955& 5.7	&2.014&7.4	&2.044& 9.1	&2.062 & 10.8 & 2.075\\
1.05 &0.906 & 2.4	&1.805&   4.1	&1.961& 5.8     &2.016& 7.5	&2.045& 9.2	&2.063 & 10.9 & 2.075\\
1.10 &1.051 & 2.5	&1.821&   4.2	&1.965& 5.9	&2.019& 7.6	&2.047& 9.3	&2.064 & 11.0 & 2.076\\
1.15 &1.153 & 2.6	&1.836&   4.3	&1.969& 6.0	&2.021& 7.7	&2.048& 9.4	&2.065 & 11.1 & 2.076\\
1.20 &1.232 & 2.7	&1.849&   4.4	&1.974& 6.1	&2.023& 7.8	&2.049& 9.5	&2.066 & 11.2 & 2.077\\
1.25 &1.293 & 2.8	&1.862&   4.5	&1.978& 6.2     &2.025& 7.9	&2.050& 9.6	&2.066 & 11.3 & 2.077\\
1.30 &1.352 & 2.9	&1.873&   4.6	&1.981& 6.3	&2.027& 8.0	&2.052& 9.7	&2.067 & 11.4 & 2.078\\
1.35 &1.400 & 3.0	&1.883&   4.7	&1.985& 6.4	&2.029& 8.1	&2.053& 9.8	&2.068 & 11.5 & 2.079\\
1.40 &1.441 & 3.1	&1.893&   4.8	&1.989& 6.5	&2.030& 8.2	&2.054& 9.9	&2.069 & 11.6 & 2.079\\
\hline                                  
\end{tabular}
\end{table}

\end{document}